\documentclass[]{JHEP3}
\usepackage{graphicx}
\usepackage{dcolumn}
\usepackage{bm,bbm,epsfig,multicol}
\usepackage{amssymb}
\usepackage{amsmath}

\newcommand{\as}{\alpha_s}

\title{Nonperturbative Effect in Threshold Resummation}
\author{Chong Sheng Li\\
Department of Physics and State Key Laboratory of Nuclear Physics and Technology, Peking University, Beijing 100871, China\\
E-mail: \email{csli@pku.edu.cn}}
\author{Zhao Li\\
Department of Physics and State Key Laboratory of Nuclear Physics and Technology, Peking University, Beijing 100871, China\\
E-mail: \email{zhli.phy@pku.edu.cn}}
\author{C.-P. Yuan\\
Department of Physics and Astronomy, Michigan State University, East Lansing, MI 48824, USA\\
E-mail: \email{yuan@pa.msu.edu}}
\preprint{\hepph{0903.1798}} \abstract{ We show that the
conventional threshold resummation calculation cannot describe well
the low energy Drell-Yan (DY) data without including the
non-perturbative correction terms which are deduced from analyzing
the asymptotic behavior of the resummation formalism. It is
demonstrated that the non-perturbative correction is generally small
for the large invariant mass DY pairs produced at the Tevatron and
the LHC. 
}


\begin{document}

\section{Introduction}
With ample data accumulated at the Fermilab Tevatron and expected at
the up-coming CERN Large Hadron Collider (LHC), precision
measurements at hadron colliders become possible. To test the
Standard Model (SM) predictions and to probe possible new physics
signatures, many higher order theoretical calculations have been
performed in recent years to study the phenomenology of
Drell-Yan (DY) pair, $W$ boson, top quark pair and Higgs boson
produced in hadron collisions. It is well known that in perturbative
QCD (pQCD) hadronic cross sections could receive logarithmically
enhanced contribution in the partonic threshold region, which
corresponds to $z\to1$, where $z=Q^{2}/\hat{s}$ with $Q$ being the
invariant mass of the final state (say, the DY pair) and
$\sqrt{\hat{s}}$ the center-of-mass (CM) energy of the partonic
process. These contributions take the form of ``plus''
distributions, such as $[\ln^{n}(1-z)/(1-z)]_{+}$, and can be
resummed to all orders in the expansion of the strong coupling
$\alpha_{s}$. This is called the threshold resummation calculation
which plays an important role for precision tests at hadron colliders.

A typical Sudakov exponent in threshold resummation
\cite{Laenen:1998qw} in the modified minimal subtraction
($\overline{\text{MS}}$) \cite{Fanchiotti:1992tu,Kidonakis:1999ze} scheme,
\begin{equation}
\begin{split}
E^{f_i}(N)=&-\int_{0}^{1}dz\frac{z^{N-1}-1}{1-z}\Big\{
\int_{(1-z)^{2}}^{1}\frac{d\lambda}{\lambda}A^{(f_i)}[\as(\lambda Q^2)]
+\nu^{(f_i)}[\as((1-z)^{2}Q^2)] 
\Big\}
\end{split}\label{efunc}\, ,
\end{equation}
generally involves singularities, known as the Landau pole of QCD
running coupling. This occurs when the relevant energy scale is
smaller than the QCD scale $\Lambda_\text{QCD}$, as $z\to 1$ or
$\lambda\to 0$, where the pQCD fails and the non-perturbative QCD
effects must set in. To avoid directly confronting the
non-perturbative contribution, 
a few approaches have been proposed in the literature.
One approach is to apply the principal value resummation \cite{Contopanagos:1993yq} 
to choose a contour to evaluate the above integral without hitting Landau
pole~\cite{Laenen:1998qw,Contopanagos:1993yq,Catani:1996yz}. 
Another method to treat the Landau pole problem was recently proposed by Bonvini 
{\it et al.} in Ref.~\cite{Bonvini:2008ei}.
Other approaches were also proposed in the literature. 
For example, in Ref.~\cite{Becher:2006nr}, 
Becher {\it et al.}  established an approach to resum the logarithmic contributions 
directly in the momentum space based on effective theory.
In Ref.~\cite{Kidonakis:2001nj}, Kidonakis {\it et al.} proposed
to approximate the resummation calculation by an expansion of 
the resummed cross section, which will be further discussed below. 
In this paper, we propose a new approach to add non-perturbative correction
terms to the minimal prescription threshold resummation formalism
\cite{Catani:1996yz}, as suggested by the perturbative expansion of joint resummation.

This paper is organized as follows. In Section 2 we shall first briefly review
the relevant part of threshold resummation (RES) formalism. In Section 3 from
the threshold resummation formalism we induce the functional form of the non-perturbative (NP)
correction terms to be added to the minimal prescription threshold
resummation calculation. Then, in Section 4 we apply this new threshold
resummation formalism, denoted here as RES+NP, to three DY
experiments with CM energy of the hadron colliders ranging from
38.76 GeV to 1960 GeV~\cite{Moreno:1990sf,Webb:2003ps,Han:2007zzd}.
From the comparison to data, we determine the coefficients in the
NP correction terms of the threshold resummation formalism. We also
compare the RES+NP result with the predictions of several popular approaches. Finally, we give our
prediction for the DY pair production at the LHC. 
Section 5 contains a brief summary of the conclusions.

\section{Threshold Resummation}

The differential cross section of the DY pair,
in terms of its invariant mass ($Q$) and rapidity ($Y$),
in the hadron CM frame (with energy $\sqrt{S}$)
is~\cite{Choudhury:2005eu}
\begin{equation}
\begin{split}
S\frac{d\sigma^{H_1H_2}}{dQ^2dY}(\tau, Y)=&
\int_0^1 dz
\int^1_{0}\frac{dx_1}{x_1}\int^1_{0}\frac{dx_2}{x_2}
f_{p_1}^{H_1}(x_1,\mu_F)
f_{p_2}^{H_2}(x_2,\mu_F)
\delta(z-\frac{\tau}{x_1x_2})
\frac{d\sigma^{p_1p_2}}{dzdy}(z,y),\label{conveq}
\end{split}
\end{equation}
where $\tau={Q^2}/{S}$.
$y$ is the rapidity of lepton pair
in the parton CM frame.
$f_{p_i}^{H_i}$ are the parton distribution functions (PDFs),
and $\mu_F$ is the factorization scale.

Conventionally, the rapidity-integrated cross sections only need the
Mellin transform to turn the convolution into multiplication.
However, for the case of the rapidity distribution the Fourier
transform with respect to $Y$ is also needed
\cite{Mukherjee:2006uu,Bolzoni:2006ky}. Applying the Mellin-Fourier
transform to Eq.(\ref{conveq}), we obtain
\begin{equation}
\begin{split}
\widetilde\omega^{H_1H_2}&\equiv\int_0^1d\tau
\tau^{N-1}\int_{\ln\sqrt{\tau}}^{-\ln\sqrt{\tau}}dYe^{iMY}S\frac{d\sigma^{H_1H_2}}{dQ^2dY}(\tau,
Y)
\\&
=\widetilde f_{p_1}^{H_1}(N+iM/2)\widetilde
f_{p_2}^{H_2}(N-iM/2)\widetilde\omega^{p_1p_2},
\end{split}
\label{omega}
\end{equation}
where 
\begin{equation}
\widetilde f_{p_i}^{H_i}(N\pm iM/2)=\int_0^1dxx^{N-1\pm iM/2}f_{p_i}^{H_i}(x_i)
\end{equation}
and
\begin{equation}
\widetilde \omega^{p_1p_2}(N,M)=\int_0^1dzz^{N-1}\int_{\ln\sqrt{z}}^{-\ln\sqrt{z}}dy e^{iMy}\frac{d\sigma^{p_1p_2}}{dzdy}
\end{equation}
are the
PDFs and differential cross section in moment space, respectively.
As shown in Ref. \cite{Mukherjee:2006uu}, near threshold region the
$M$ dependence in $\widetilde\omega^{p_1p_2}(N,M)$ is negligible,
and it is a good approximation to take the well-known resummed form
\cite{Kidonakis:1999ze,Mukherjee:2006uu} for the rapidity-integrated cross section,
\begin{equation}
\begin{split}
\widetilde\omega^{p_1p_2}_\text{res}(N,M)&\approx\widetilde\omega^{p_1p_2}_\text{res}(N)=\exp\Big[\sum_i E^{p_i}(N)\Big]
\exp\Big[\sum_i 2\int_{\mu_F}^{Q}\frac{d\mu}{\mu}\gamma_{p_i/p_i}(\as(\mu^2))\Big]
\\&
\times \text{Tr}\Big\{H^{p_1p_2}(\as(\mu_R^2))\bar{P}\exp\Big[\int_Q^{Q/\tilde N}\frac{d\mu}{\mu}\Gamma^{\dagger p_1p_2}_S(\as(\mu^2))\Big]
\\&
\times\tilde S^{p_1p_2}\Big(1,\as\Big(\frac{Q^2}{\tilde N^2}\Big)\Big)P\exp\Big[\int_Q^{Q/\tilde N}\frac{d\mu}{\mu}\Gamma_S^{p_1p_2}(\as(\mu^2))\Big]\Big\},
\label{rescs}
\end{split}
\end{equation}
where $\tilde{N}=N\exp{(\gamma_E)}$, $\gamma_E$ is the Euler constant, and $\mu_R$ is the renormalization scale.
$P$ denotes path ordering \cite{Collins:1981uw}.
The first exponent in Eq.(\ref{rescs}), as given in Eq. (\ref{efunc}), resums the collinear and soft gluon emission contributions from the initial state, with
\begin{equation}
A^{(f_i)}(\as)=C_f\Big(\frac{\as}{\pi}+\frac{1}{2}K\Big(\frac{\as}{\pi}\Big)^2\Big),
\end{equation}
\begin{equation}
K=C_{A}\Big(\frac{67}{18}-\frac{\pi^2}{6}\Big)-\frac{5}{9}n_{f},
\end{equation}
\begin{equation}
\nu^{(f_i)}(\as)=\frac{2C_f}{\pi}\as,
\end{equation}
where $n_f$ is the flavor number of light quarks. $C_f=C_F=(N_c^2-1)/(2N_c)$ for initial state quarks and $C_f=C_A=N_c$ for initial state gluons. $N_c$ is the number of colors.
In Eq.(\ref{rescs}), at the one-loop order,
$\gamma_{q/q}=(\as/\pi)(\frac{3}{4}C_{F}-C_{F}\ln\tilde{N})$ for quarks,
and $\gamma_{g/g}=(\as/\pi)(\beta_0-C_A\ln\tilde{N})$ for gluons.
The $\beta$ function is defined as
\begin{equation}
\beta(\as)=\frac{1}{2}\mu\frac{d\ln g_s}{d\mu}=-\sum_{n=0}^{\infty}\beta_n\Big(\frac{\as}{\pi}\Big)^{(n+2)},
\end{equation}
with
\begin{equation}
\beta_0=(11C_A-2n_f)/12,
\end{equation} 
\begin{equation}
\beta_1=(17C_A^2-5C_An_f-3C_Fn_f)/24.
\end{equation}
$\Gamma_S$ in Eq.(\ref{rescs}) is the soft anomalous dimension matrix, which can be derived from the eikonal diagrams \cite{Laenen:1998qw,Kidonakis:1999ze} and is given at the one-loop order by
\begin{equation}
\Gamma_{S}=\frac{\as C_{F}}{\pi}(1-\pi i).
 \end{equation}
At the next-to-leading logarithm (NLL) accuracy, $\widetilde
\omega^{p_1p_2}$ is approximated as \cite{Mukherjee:2006uu}
\begin{equation}
\widetilde\omega^{p_1p_2}_\text{NLL}(N)=\widetilde\omega^{p_1p_2}_BC(\as)\exp[g_1(\lambda)\ln \tilde{N}+g_2(\lambda)],
\label{sfunc}
\end{equation}
with
\begin{equation}
\lambda=\beta_0\alpha_s\ln\tilde{N}/\pi,
\end{equation}
\begin{equation}
g_1(\lambda)=\frac{C_F}{\beta_0\lambda}\Big[2\lambda+(1-2\lambda)\ln(1-2\lambda)\Big],
\end{equation}
\begin{equation}
\begin{split}
g_2(\lambda)&=\frac{C_F\beta_1}{\beta_0^3}\Big[2\lambda+\ln(1-2\lambda)+\frac{1}{2}\ln^2(1-2\lambda)\Big]
-\frac{C_FK}{2\beta_0^2}[2\lambda+\ln(1-2\lambda)]
\\&
+\frac{C_F}{\beta_0}[2\lambda+\ln(1-2\lambda)]\ln\frac{Q^2}{\mu_R^2}-\frac{C_F}{\beta_0}2\lambda\ln\frac{Q^2}{\mu_F^2},
\end{split}
\end{equation}
where $\widetilde\omega^{p_1p_2}_B$ is the Born differential cross section in moment space.
By matching the moments of the NLO cross section, the coefficient $C(\as)$ in Eq.(\ref{sfunc}) can be obtained as follows \cite{Mukherjee:2006uu}
\begin{equation}
C(\as)=1+\frac{\as}{\pi}C_F\Big(-4+\frac{2\pi^2}{3}+\frac{3}{2}\ln\frac{Q^2}{\mu_F^2}\Big).
\end{equation}
The cross section in momentum space can be obtained via inverse Mellin-Fourier transform
\begin{equation}
\omega^{p_1p_2}(z, y)=
\frac{1}{2\pi i}\int_{C-i\infty}^{C+i\infty} dN z^{-N}
\frac{1}{2\pi}\int_{-\infty}^{\infty} dMe^{-iMy} \widetilde\omega^{p_1p_2}(N,M).
\label{invtr}
\end{equation}
In order to avoid double-counting the fixed order contributions 
up to the next-to-leading order (NLO),
it is needed to 
subtract the first two orders of $\as$ 
expansion to obtain the final cross section, i.e.,
\begin{equation}
\begin{split}
S\frac{d\sigma^{H_1H_2}_\text{RES}}{dQ^2dY}&=S\frac{d\sigma^{H_1H_2}_\text{NLO}}{dQ^2dY}+S\frac{d\sigma^{H_1H_2}_\text{NLL}}{dQ^2dY}
-\Big(S\frac{d\sigma^{H_1H_2}_\text{NLL}}{dQ^2dY}\Big)_{\as=0}
-\as\Big(\frac{\partial}{\partial\as}S\frac{d\sigma^{H_1H_2}_\text{NLL}}{dQ^2dY}\Big)_{\as=0} \,,
\end{split}
\end{equation}
where 
$\sigma^{H_1H_2}_\text{NLO}$,  
corresponding to  $\omega^{p_1p_2}_\text{NLO}$ in the momentum space,  
denotes the NLO cross section. 

For comparison, in this paper we also investigate the next-to-next-to-leading order (NNLO) expansion \cite{Kidonakis:2001nj,Kidonakis:2003tx}
of the resummed cross sections. 
At first, in order to reproduce the fixed order expressions, we recover the $M$ dependence in 
\begin{equation}
\widetilde\omega^{p_1p_2}(N,M)=\frac{1}{2}[\widetilde\omega^{p_1p_2}(N+iM/2)+\widetilde\omega^{p_1p_2}(N-iM/2)],
\end{equation}
although we have neglected the $M$ dependence in the numerical evaluation of 
the resummed cross section \cite{Mukherjee:2006uu}.
For the inverse Mellin-Fourier transform of the logarithms $\ln^k(N\pm iM/2)$ we find
\begin{align}
I_0=&\frac{1}{2\pi i}\int_{C-i\infty}^{C+i\infty} dN z^{-N}
\frac{1}{2\pi}\int_{-\infty}^{\infty} dMe^{-iMy}\nonumber
\\
=&\delta(1-z)\delta(y)\nonumber
\\
=&\delta(1-z)\delta(y\pm \frac{1}{2}\ln z)\nonumber
\\
=&\frac{1}{2\pi i}\int_{C-i\infty}^{C+i\infty} dN z^{-N}\delta(y\pm \frac{1}{2}\ln z),
\\
I_k=&\frac{1}{2\pi i}\int_{C-i\infty}^{C+i\infty} dN z^{-N}
\frac{1}{2\pi}\int_{-\infty}^{\infty} dMe^{-iMy}[\ln^k(N\pm iM/2)+\gamma_E]\nonumber
\\
=&\frac{1}{2\pi i}\int_{C-i\infty}^{C+i\infty} dN' z^{-N'}
\frac{1}{2\pi}\int_{-\infty}^{\infty} dMe^{-iM(y\pm \frac{1}{2}\ln z)}[\ln^k(N')+\gamma_E]\nonumber
\\
=&\frac{1}{2\pi i}\int_{C-i\infty}^{C+i\infty} dN z^{-N}\ln^k(\tilde{N})\delta(y\pm\frac{1}{2}\ln z).
\end{align}
Hence, we can replace the inverse Fourier transform and $M$ dependence by the function $[\delta(y+\frac{1}{2}\ln z)+\delta(y-\frac{1}{2}\ln z)]/2$.
The remaining calculation is the same as the rapidity-integrated differential cross section.
If we expand the exponent of the resummed cross section in Eq.(\ref{sfunc}) to the NLO accuracy \cite{Kidonakis:2003tx}, the differential cross section for DY assuming $\mu_F=\mu_R=Q$ is
\begin{equation}
\omega^{p_1p_2(1)}_\text{exp.}(z,y)=\omega^{p_1p_2}_0\frac{\as}{\pi}[c_3 D_1(z)+c_1\delta(1-z)]\frac{\delta(y+\frac{1}{2}\ln z)+\delta(y-\frac{1}{2}\ln z)}{2},
\end{equation}
where $D_k(z)=[\ln^k(1-z)/(1-z)]_+$, $c_3=4C_F$ and $c_1=2C_F\zeta_2-4C_F$.
$\omega^{p_1p_2}_0$ is the coefficient of $\delta(1-z)\delta(y)$ in the Born differential cross section $\omega^{p_1p_2}_B$. 
It is evident that this reproduces the dominant contribution near threshold in the NLO differential cross section, as given in Ref. \cite{Anastasiou:2003yy}.
Similarly, at the NNLO accuracy, the expansion of the exponent in Eq. (\ref{sfunc}) yields 
\begin{equation}
\begin{split}
\omega^{p_1p_2(2)}_\text{exp.}(z,y)=&\omega^{p_1p_2}_0\Big(\frac{\as}{\pi}\Big)^2
\Big\{\frac{1}{2}c_3^2D_3(z)-\beta_0 c_3 D_2(z)
+(c_3c_1-\zeta_2c_3^2+2C_FK)D_1(z)
\\
&+(\zeta_3c_3^2-2\beta_0 c_1)D_0(z)
+(\frac{1}{2}c_1^2+\frac{1}{4}\zeta_2^2c_3^2-\frac{3}{4}\zeta_4c_3^2)\delta(1-z)\Big\}
\\
&\times
\frac{\delta(y+\frac{1}{2}\ln z)+\delta(y-\frac{1}{2}\ln z)}{2}.
\end{split}
\end{equation}
Below, we define the NNLO-NLL (next-to-next-to-leading order and next-to-leading logarithmic) \cite{Kidonakis:2004ib} corrected differential cross section as
\begin{equation}
\omega^{p_1p_2}_\text{NNLO-NLL}=\omega^{p_1p_2}_\text{NLO}+\omega^{p_1p_2(2)}_\text{exp.}.
\end{equation}

\section{Nonperturbative Effect}
 While deriving Eq.(\ref{sfunc}) from Eq.(\ref{efunc}),
approximation \cite{Catani:1996yz} has been made to avoid Landau
pole in the original expansion, and the integration in
Eq.(\ref{efunc}) was carried out via perturbative expansion. This
approximation could fail if the NP contribution in Eq.(\ref{efunc})
is large. Hence, we propose to add NP correction terms in the
resummation formalism to better approximate the total contribution
from Eq.(\ref{efunc}). To find out the proper functional form to
parameterize the NP corrections, we examine the joint resummation
formalism in Ref. \cite{Laenen:2000ij}, where the resummed cross
section in moment (and impact parameter) space is expressed as
\begin{equation}
\hat{\sigma}_{ab}^{(eik)}(N,b)=\exp[D_{ab}^{(eik)}(N,b)]\exp[E_{ab}^{(eik)}(N,b)],
\end{equation}
where
\begin{equation}
\begin{split}
E_{ab}^{(eik)}(N,b)=&\int_0^{Q^2}\frac{dk_T^2}{k_T^2}\Big\{
\sum_{i=a,b} A_i(\as(k_T))
\Big[J_0(bk_T)
K_0\Big(\frac{2Nk_T}{Q}\Big)+\ln\Big(\frac{\tilde{N}k_T}{Q}\Big)\Big]\Big\}
\\&
-\ln\tilde{N}\int_{\mu_F^2}^{Q^2}\frac{dk_T^2}{k_T^2}\sum_{i=a,b}A_i(\as(k_T)) ,
\end{split}
\end{equation}
\begin{equation}
\begin{split}
D_{ab}^{(eik)}(N,b)=&\int_0^{Q^2}dk_T^2{\mathcal A}_{ab}(\as(k_T),k_T)
\Big[\ln\Big(\frac{k_T}{Q}\Big)
+e^{-ib\cdot k_T}K_0\Big(\frac{2Nk_T}{Q}\Big)\Big]
\\&+\int_0^{Q^2}dk_T^2
\int_0^{Q^2-k_T^2}dk^2 w_{ab}(k^2,k_T^2,\as(\mu_F^2))
\\&
\times
\Big[e^{-ib\cdot k_T}\Big\{K_0\Big(2N\sqrt{\frac{k_T^2+k^2}{Q^2}}\Big)
-K_0\Big(\frac{2Nk_T}{Q}\Big)\Big\}
+\ln\Big(\sqrt{\frac{k_T^2+k^2}{k_T^2}}\Big)\Big].
\end{split}
\end{equation}
The infrared renormalon singularities occur as $k_T\to 0$ for a
large $Q$ value with $Q \gg N k_T$. Below, we shall ignore the
$b$-dependent term which is only relevant to transverse momentum
resummation. From the expansion of the Bessel function $K_0(x)$ for
small value of argument $x$, $K_0(x)\sim -\ln(x e^{\gamma_E/2})$, we
find that the leading behavior of $E_{ab}^{(eik)}$  in $k_T\to 0$
limit can be described by the following two functions: 
\begin{equation}
\frac{N^2}{Q^2} \quad \text{and} \quad \frac{N^2}{Q^2}\ln\Big(\frac{Q}{\tilde{N}Q_0}\Big).
\end{equation} 
Since ${\mathcal A}_{ab}$ behaves as
$1/Q^{2}$  when $k_T\to 0$ with a large $Q$
value~\cite{Laenen:2000ij},
 $D_{ab}^{(eik)}$ behaves as $1/Q^4$ which is suppressed as compared to
$E_{ab}^{(eik)}$.
Another correction term at the order of $1/Q^2$, but suppressed by $1/N^2$,
can be obtained by examining the behavior of $E_{ab}^{(eik)}$ and
 $D_{ab}^{(eik)}$  in the limit that  $k_T\to 0$ and $Q$ is not very
large as compared to $k_T$, i.e., $Q \sim N k_T$.
Since for large argument $x$, $K_0(x)\sim e^{-x}/\sqrt{x}$, we
can neglect the contribution from $K_0$.
The leading contribution in this limit comes from
 the logarithm term $\ln(k_T/Q)$ in
$D_{ab}^{(eik)}$ which implies that
\begin{equation}
\frac{1}{Q^2}\ln\Big(\frac{Q}{Q_0}\Big)
\end{equation} 
should be considered.

After combining the above three sources, we obtain the
NP correction term
\begin{equation}
{\cal
S}^\text{NP}(N)=\frac{N^2}{Q^2}\Big(a_1+a_2\ln\frac{Q}{\tilde{N}Q_0}\Big)+a_3\frac{1}{Q^2}\ln\frac{Q}{Q_0}\, ,
\label{NP1}
\end{equation}
where the parameter $Q_0$ is chosen to be $1.3$~GeV\footnote{This is the energy scale that the CTEQ6.6 parton distribution functions are evolved from.}
and the dimensional NP parameters $a_1$, $a_2$ and $a_3$ are to be
determined by comparing the theory prediction of RES+NP with
experimental data. 
In this improved threshold resummation formalism,
$\widetilde\omega^{p_1p_2}$ in Eq.(\ref{omega}) is written in the
form
\begin{equation}
\tilde{\omega}_\text{res+NP}^{p_1p_2}(N)=
\tilde{\omega}_\text{res}^{p_1p_2}(N)\exp[{\cal S}^\text{NP}(N)].
\label{NP2}
\end{equation}

\section{Numerical Results}

In our numerical calculation, the SM parameters are chosen as in Ref. \cite{Amsler:2008zz}.
The running QCD coupling is evaluated at the three-loop order \cite{Amsler:2008zz},
and CTEQ6.6M PDFs \cite{Nadolsky:2008zw} are used with $\mu_F=\mu_R=Q$.
In Table \ref{datasets}, we list the set of DY experimental data
to be considered in our analysis.
 \TABULAR[h]{|c|c|c|c|}{\hline
 Experiment & no. of data points & $\sqrt{S}$(GeV) &
$\sigma_n^N$
\\
\hline
E605 \cite{Moreno:1990sf} & 119 & 38.76 & 15\%
\\
E866 ($pp$) \cite{Webb:2003ps} & 184 & 38.76 & 6.5\%
\\
CDF \cite{Han:2007zzd} & 29 & 1960 & 5\%\\
\hline}{The DY data sets considered in the analysis.
$\sigma_n^N$ is the experimental normalization uncertainty of the $n$th experiment.\label{datasets}}
We follow the method of least chi-square ($\chi^2$) analysis
in Ref.~\cite{Stump:2001gu} to find the best fit by allowing the
overall normalization of each experiment to vary.
They are denoted as
 $N_1$, $N_2$ and $N_3$ for E605, E866 and CDF Run-2 DY
 (via $Z/\gamma^\ast$ production) experiments, respectively.
The NP parameters are determined by the
best fit of the RES+NP calculation to the set of DY data, which yields
\begin{equation}
a_1=-0.60~\text{GeV}^2, \quad a_2=-2.87~\text{GeV}^2,
\quad a_3=5.58~\text{GeV}^2, \nonumber \label{aivalue}
\end{equation}
with $N_i$ being around 1 and $\chi^2$ per degree of freedom (dof)
1.07, cf. Table \ref{chisq}.

In Table \ref{chisq} , we also compare the result of RES+NP
calculation with a few other theory calculations which include the
NLO \cite{Choudhury:2005eu,Anastasiou:2003yy,Mathews:2004xp},
NNLO \cite{Melnikov:2006kv}, NNLO-NLL \cite{Kidonakis:2003tx}, and the
usual threshold resummation calculation (RES)
\cite{Mukherjee:2006uu}. 
.
We repeat the same fitting procedure for
each theory calculation by allowing the normalization of each data
set to float within its experimental uncertainty in order to find
the best fit to theory prediction.

\TABULAR[h]{|c|c|c|c|c|c|}{\hline
&  $\chi^2_\text{E605} (N_1)$ & $\chi^2_\text{E866} (N_2)$
&$\chi^2_\text{CDF} (N_3)$ &  $\chi^2_\text{total}$ &
$\chi^2/\text{dof}$
\\
\hline
NLO & 103 (0.99) & 247 (0.97) & 19.5 (0.95) & 369.5 & 1.11
\\
\hline
NNLO &  271 (1.25) & 772 (1.28) & 20.6 (0.92) & 1063.6 & 3.20
\\
\hline
NNLO-NLL & 147 (1.04) & 908 (1.05) & 17.4 (0.94) & 1072.4 & 3.23
\\
\hline
RES & 196 (1.18) & 1198 (1.20) & 17.6 (0.97) & 1411.6 & 4.25
\\
\hline
RES+NP & 128 (1.03) & 209 (0.96) & 17.0 (0.97)  & 354.0 & 1.07
\\
\hline}{The minimal value of $\chi^2$ and normalization factor for
each experiment and theory prediction.\label{chisq}}
As shown in the table, the NLO results agree well with the data
because the E605 data were included in determining the CTEQ6.6M PDFs
at the NLO. The NNLO result is about a factor 3 worse than the NLO
result in $\chi^2/\text{dof}$, owing to the low energy E605 and E866
data, which indicates that a NNLO PDF set is needed to improve the
comparison. 
The NNLO-NLL prediction is simliar to the RES results.
The conventional threshold resummation calculation (RES) cannot describe
well the E605 and E866 data with the mass of the DY pair ranging
from 7 GeV to 18 GeV and 4.2 GeV to around 15 GeV, respectively. The
largest difference between the results of RES and RES+NP occurs in
the E866 data. To examine it in more detail, we show in
Fig.\ref{chidist} the comparison among various theory calculations
for one particular set of E866 data, with $0.60<x_F<0.65$, as an
example, where $x_F$ is the Feynman-$x$ variable \cite{Webb:2003ps}.
To clearly examine the difference between the theoretical predictions
and the experimental data, we define
$\chi=(D_n-T_{n}/N_{2})/\sigma_{n}^D$, where $D_{n}$, $\sigma_{n}^D$
and $T_{n}$ denote the data value, experimental measurement
uncertainty and the theoretical value for the $n$th data point,
respectively. As shown in the figure, the largest deviation from
data occurs when $Q$ is around 5 to 6 GeV.
The RES result also fails to describe data unless the NP
correction terms are included which is the result of RES+NP. Hence,
we conclude that to describe the low energy DY data with the
threshold resummation formalism, the NP correction terms must be
included in order to take into account the part of contribution missing
from approximating the Sudakov integral Eq.(\ref{efunc}) by its
perturbative expansion Eq.(\ref{sfunc}).

\EPSFIGURE[h]{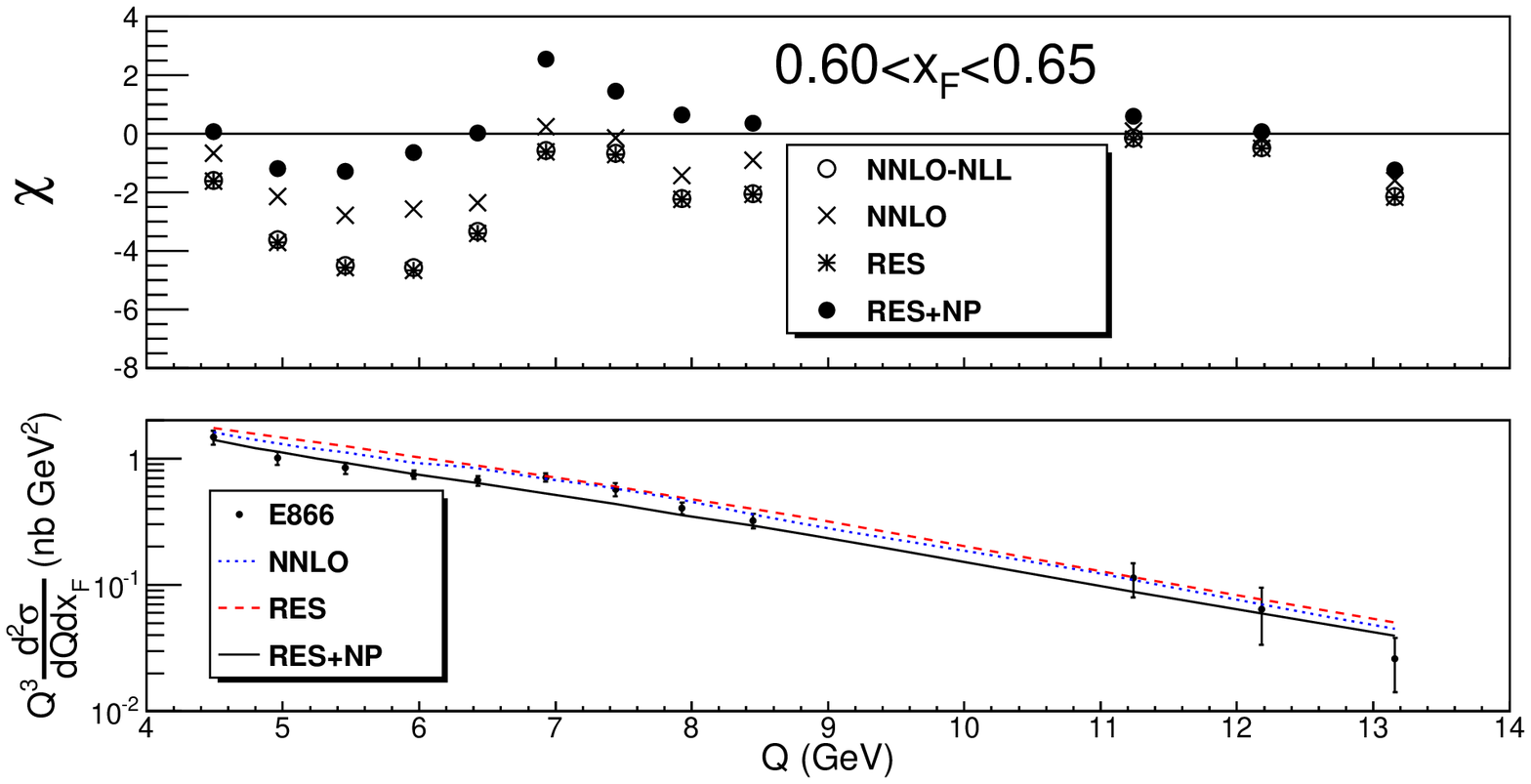,width=\textwidth} {Different theoretical
predictions compared to the E866 ($pp$) experimental data, with
$0.60<x_F<0.65$. In the lower plot, the theoretical results have
been multiplied by $1/N_2$. \label{chidist}}

Next, let us compare various theory calculations to the large $Q$
(around 100 GeV) DY data, taken by the CDF Collaboration at the
Tevatron Run-2.

\EPSFIGURE[h]{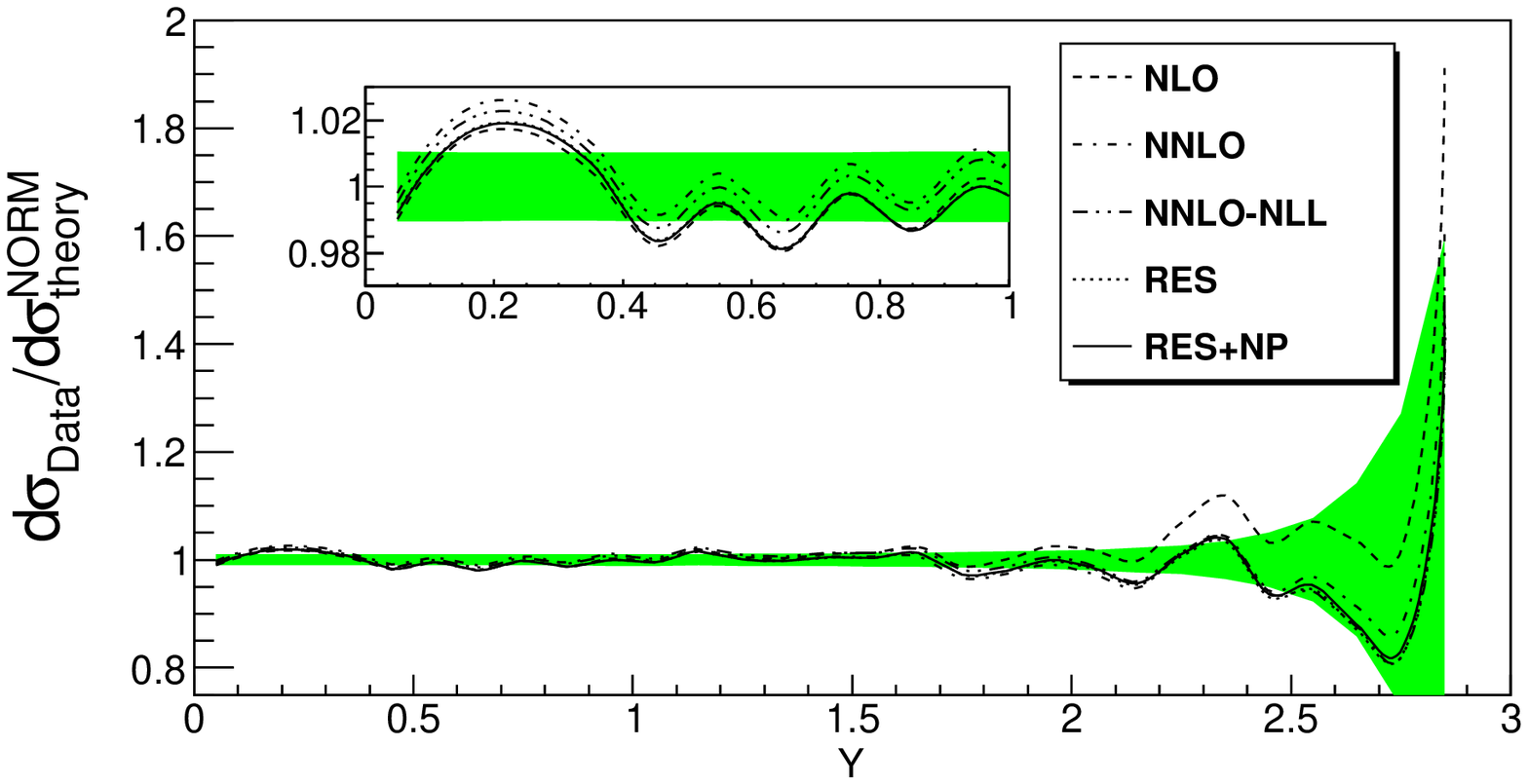,width=\textwidth} {The ratio of CDF
Run-2 DY data to various theoretical predictions (after being
multiplied by $1/N_3$). The shaded area indicates the statistical
error of the data.\label{ratioy}}

As shown in Fig. \ref{ratioy}, for $Y>2$, the NLO result becomes
smaller than the data while the NNLO, RES and RES+NP calculations
give similar results, though the RES+NP result gives the lowest
$\chi^2$ for this set of data, cf. Table \ref{chisq}. 
We could also compare the total cross section of the DY
pair produced at the Tevatron Run-2 and the LHC. The result of
comparison is listed in Table \ref{totalcs}.
\TABULAR[h]{|c|c|c|c|c|}{\hline
 $\sigma (p\bar p/pp\to l^+l^-+X) (\text{pb}) $ & NLO & NNLO &
RES & RES+NP
\\
\hline
Tevatron & 240.7 & 242.0 & 248.0 & 247.98
\\
\hline
LHC & 2047.9 & 2036.8 & 2115.2 & 2115.3
\\
\hline}{The total cross sections of the DY pair (with
$66 \/ \text{GeV}<Q< 116 \/ \text{GeV}$) produced at the Tevatron Run-2 and the LHC.\label{totalcs}}
It shows that the results of RES and RES+NP are about the same, and
differ from NNLO (and NLO) by about $2.5\%$ at the Tevatron and
$3.7\%$ at the LHC. Hence, we conclude that the effect from the NP
correction terms to the conventional threshold resummation formalism
for large $Q$ values in high energy hadron collisions is not
important.
\begin{figure}[h]
\caption{\label{lhcdist}
The differential cross section as a function of the invariant mass of
the DY pair produced at the LHC. }
\includegraphics[width=\textwidth]{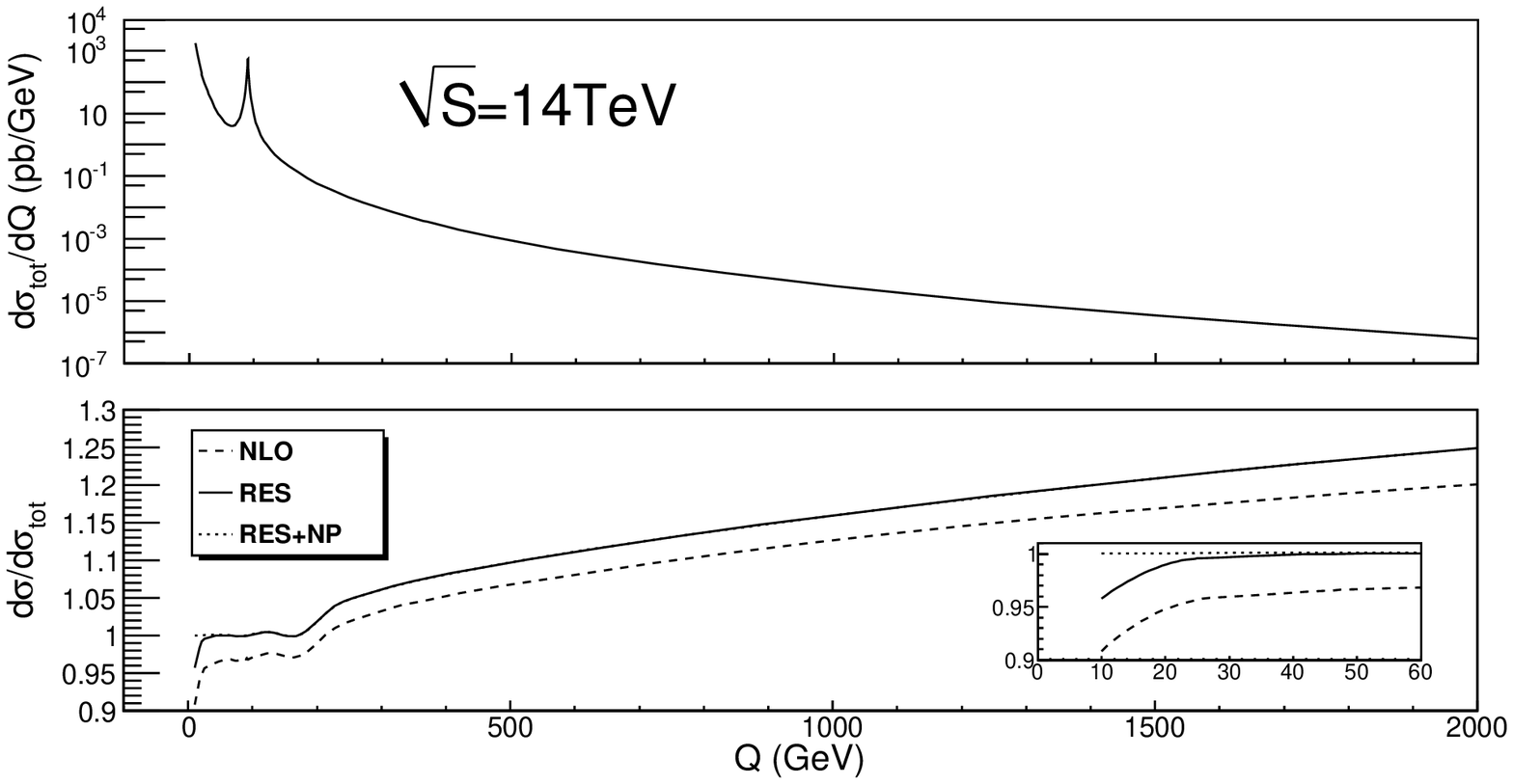}
\end{figure}
Finally, we show in Fig. \ref{lhcdist} the differential cross
section as a function of the invariant mass of the DY pair produced
at the LHC. In this figure, $d\sigma_\text{SUM}/dQ$ represents the
differential cross section including the RES+NP result and the
electroweak box diagram contribution, originated from the
re-scattering of $W$ and $Z$ bosons in loop diagrams~\cite{Baur:2001ze}. In
order to study in detail the higher order correction to the shape of
the $d \sigma/ dQ$ distribution, we also plot the ratio of
differential cross sections normalized by $d\sigma_\text{SUM}/dQ$. The
label RES+NP indicates the ratio of
$d\sigma_\text{RES+NP}/d\sigma_\text{SUM} $, {\it etc.} It is
evident that the results of RES and RES+NP are almost the same
except when the value of $Q$ is small, less than about 30 GeV.

\section{Conclusion}
In conclusion, we have investigated the NP effect in the threshold
resummation calculations. From analyzing the asymptotic behavior of
the conventional threshold resummation formalism (RES), we proposed
the NP correction terms to be included in the improved threshold
resummation formalism (RES+NP). They are subsequently determined by
fitting to the DY data from E605, E866 ($pp$) and CDF Run-2
experiments. We found that to describe the low energy DY data with
relatively small value of invariant mass, the NP effect cannot be
ignored in threshold resummation calculation. 
In contrast, the minimal prescription threshold resummation
formalism (RES) gives a similar prediction as RES+NP when the
invariant mass of the DY pair is large. Though it remains to be seen
how well the threshold resummation formalism could describe the top
quark pair production rates at the Tevatron and the
LHC~\cite{Czakon:2008cx}, the type of NP effect discussed in this
paper is not expected to be important. On the contrary, this effect
will become relevant for describing the low invariant mass DY pairs
produced at the Relativistic Heavy Ion Collider (RHIC) at the
Brookhaven Laboratory.

\begin{acknowledgments}
We thank N. Kidonakis for a useful communication. 
This work was supported in part by the National Natural Science
Foundation of China, under Grants No. 10721063 and No. 10635030. CPY
was supported in part by the U.S. National Science Foundation under
Grant No. PHY-0555545.
\end{acknowledgments}

\end{document}